\documentclass[aps,prl,reprint,groupedaddress,showpacs,preprintnumbers]{revtex4-1}
\PassOptionsToPackage{caption=false}{subfig}
\usepackage{subfig} 
\usepackage{subfloat}
\usepackage{mathrsfs}
\usepackage{graphicx}
\usepackage{xspace}
\usepackage{afterpage}
\usepackage{multirow}
\usepackage{dcolumn}
\usepackage{supertabular}
\usepackage{feynmf}
\usepackage[hyperref]{xcolor}
\usepackage[hyperindex,breaklinks]{hyperref}
\usepackage{atlasphysics}

\newcommand{\Mgj}{\ensuremath{m_{\gamma j}}\xspace}
\newcommand{\qstar}{\ensuremath{q^{*}}\xspace}
\newcommand{\Mqstar}{\ensuremath{m_{\qstar}}\xspace}

\newcommand{\pb}{\ensuremath{\text{pb}}\xspace}
\newcommand{\lumfb}{\ensuremath{\text{fb}^{-1}}\xspace}
\newcommand{\bumphunter}{{\sc BumpHunter}\xspace}
\newcommand{\pythia}{{\sc Pythia}\xspace}

\newcommand{\jetphox}{{\sc jetphox}\xspace}
\newcommand{\acceptance}{{\rm A}\xspace}
\newcommand{\branching}{{\rm B}\xspace}
\newcommand{\antikt}{anti-$k_{\rm t}$}

\begin{document}

\preprint{CERN-PH-EP-2011-203} 
\preprint{Submitted to Physical Review Letters}
\vspace*{0.2cm}
\title{Search for Production of Resonant States in the Photon+Jet Mass Distribution \\ using $pp$ Collisions at $\sqrt{s}=7\;\tev$ Collected by the ATLAS Detector}
\author{ATLAS Collaboration}

\begin{abstract}
  This Letter describes a model-independent search for production of
  new resonant states in photon+jet events in 2.11~\lumfb of
  proton-proton collisions at $\sqrt{s}=7$~\tev. We compare the
  photon+jet mass distribution to a background model derived from data
  and find consistency with the background hypothesis. Given the lack
  of evidence for a signal, we set 95\% CL limits on generic
  Gaussian-shaped signals and on a benchmark excited-quark (\qstar)
  model, excluding 2 \tev Gaussian resonances with cross section times
  branching fraction times acceptance times efficiency near $5\;\text{\rm
    fb}$ and excluding \qstar masses below 2.46 \tev, respectively.
\end{abstract}

\pacs{12.60.-i,13.85.Rm,14.70.Bh,14.80-j}
% full list of applicable pacs
% 11.25.-w strings and branes
% 12.60.-i Models beyond the standard model
% 12.60.Jv Supersymmetric models
% 12.60.Rc composite models
% 11.25.Wx String and brane phenomenology
% 13.87.Ce Production
% 13.85.Ni Inclusive production with identified hadrons
% 13.85.Qk Inclusive production with identified leptons, photons, or other nonhadronic particles
% 13.85.Rm Limits on production of particles
% 13.40.-f Electromagnetic processes and properties
% 14.70.Bh Photons
% 14.80.-j Other particles (including hypothetical)

\maketitle

In the Standard Model (SM), proton-proton ($pp$) collisions do not
produce photon+jet pairs through a resonance. Direct photon+jet
production occurs at tree level via Compton scattering of a quark and
a gluon or through quark-antiquark annihilation, the former accounting
for the majority of direct photon+jet production at all center-of-mass
energies. Events with a high transverse momentum photon and one or
more jets can also arise from radiation off final-state quarks or from
multi-jet processes where dijet or higher-order events produce
secondary photons during fragmentation of the hard-scatter quarks and
gluons
\cite{Catani:2002ny,Aurenche:2006vj,Belghobsi:2009hx,PhysRevD.82.014015}. The
photon+jet invariant mass (\Mgj) \cite{Note0} distribution resulting from this
mixture of processes is smooth and rapidly falling, constituting a
promising place to look for resonances.

Despite many possible exotic production mechanisms such as excited
quarks \cite{Bhattacharya:2009p2168,PhysRevD42815}, quirks
\cite{kang-and-luty,PhysRevD.83.035019,Okun19801}, Regge excitations
of string theory
\cite{PhysRevD.78.016005,PhysRevD.60.034004,PhysRevLett.101.241803,Nath2010185},
and topological pions \cite{Bai2010292}, the most recent searches for
photon+jet resonances were published a decade ago
\cite{PhysRevLett.72.3004,adloff-2000-17,zeus-collaboration-2002-549,PhysRevD-65-052006}. The
previous most sensitive search for new phenomena in the photon+jet
final state places limits on effective signal cross sections, cross
section times branching fraction times acceptance times efficiency, of
the order of $1$ \pb and on excited-quark masses up to 460~\gev at the
95\% confidence level \cite{PhysRevLett.72.3004}.

This Letter describes a general search for resonant $s$-channel
photon+jet production in 2.11~\lumfb of $pp$ collisions at a
center-of-mass energy $\sqrt{s} = 7$~\tev with the ATLAS detector. It
follows earlier measurements of isolated photon differential cross
sections at the Large Hadron Collider (LHC)~\cite{Aad:2011tw,Chatrchyan:2011ue}. The entire
\Mgj distribution is fit to a smooth function to obtain the background to
this search. We look for evidence of a narrow resonance, not much
wider than the detector mass resolution. This search extends the
method used in the search for resonant dijet production
\cite{Aad:2011fq} to handle the more diverse mixture of processes
contributing to the \Mgj distribution. In the absence of signal, we
use Bayes' theorem to set limits on Gaussian-shaped resonances and on a
benchmark excited-quark (\qstar) model
\cite{Bhattacharya:2009p2168,PhysRevD42815}.

In the excited-quark model studied here, the LHC could produce single
\qstar states with vector-like couplings to the \Wboson and \Zboson
gauge bosons via the absorption of a gluon by a quark. As in
Ref.~\cite{PhysRevD42815}, we define the model by one parameter, the
excited-quark mass $\Mqstar$, setting the compositeness scale equal to
$\Mqstar$ and SU(3), SU(2), and U(1) coupling multipliers $f_s = f =
f'= 1$. At $\Mqstar = 2.5$~\tev, this gives branching fractions for
$u^{*} \rightarrow u g$ and $u^{*} \rightarrow u \gamma$ of 0.85 and
0.02, respectively. The corresponding branching fractions for $d^{*}$
quarks are 0.85 and 0.005, respectively. We do not make any further
assumptions about higher-order corrections or the excited-quark
dynamics and neglect scale uncertainties and uncertainties on parton
distribution functions (PDFs) in order to provide a convenient
benchmark process for theoretical reinterpretation.

We simulate the SM direct photon processes and the \qstar model with
\pythia 6.4.25 \cite{Pythia:2001} using the AMBT1
tune~\cite{ATLAS-CONF-2010-031}, MSTW2007 PDFs~\cite{Sherstnev:2007nd},
and a GEANT4-based detector simulation
\cite{geant,mcsim}. Supplementary studies of the background shape
function are performed with next-to-leading-order \jetphox 1.3.0
\cite{Catani:2002ny,Aurenche:2006vj}. Additional
inelastic $pp$ interactions, termed pile-up, are included in the event
simulation, distributed so as to reproduce the number of collisions
per bunch crossing in the data. The mean number of pile-up
interactions is approximately 6.

A detailed description of the detector is available in
Ref.~\cite{1748-0221-3-08-S08003}. Photons are detected by a
lead/liquid-argon 
sampling electromagnetic calorimeter (EMC) with an
accordion geometry. In front of the EMC, the inner detector allows an
accurate reconstruction of tracks from the primary $pp$ collision
point and also from secondary vertices, permitting an efficient
reconstruction of photon conversions in the inner detector up to a
radius of 80 cm. For $|\eta| < 1.37$~\cite{Note1}, an
iron/scintillator-tile 
calorimeter behind the EMC provides hadronic
coverage. The endcap and forward regions, $1.5 < |\eta| < 4.9$, are
instrumented with liquid-argon calorimeters for both electromagnetic
and hadronic measurements. For further details relevant to photon
identification and measurement, see Ref.~\cite{Aad:2011tw}. For
details relevant to jet detection and measurement, see
Ref.~\cite{springerlink:10.1140/epjc/s10052-010-1512-2}. Events are
collected with a trigger requiring at least one photon candidate with
transverse momentum (\pt) above 80~\gev. The integrated luminosity of
the sample is $(2.11~\pm~0.08)\,\lumfb$
\cite{ATLAS-CONF-2011-116,Aad:2011dr}.

Events containing at least one photon and at least one jet are
selected for analysis. Each event must have a primary vertex with at
least five charged-particle tracks with $\pt > 400$ \mev. Multiple
vertices can appear when pile-up interactions occur for the same bunch
crossing. If more than one vertex is found, the primary vertex is
taken as the vertex with the highest scalar sum $\pt^2$ of associated
tracks. Photon candidates with $\pt > 85$ \gev and $|\eta|<1.37$ and
jet candidates with $\pt > 30$~\gev and $|\eta| < 2.8$ are used. These
objects are identified using criteria that closely follow those
applied in the isolated photon cross section measurement
\cite{Aad:2011tw} and dijet resonance search
\cite{Aad:2011fq}. Subleading photons or jets are allowed; when more
than one photon or jet is found, the highest \pt candidates are
selected to constitute the photon+jet pair.

Jets are reconstructed from clusters of calorimeter cells using the
\antikt~clustering algorithm \cite{Cacciari:2008gp} with radius
parameter $R = 0.6$. Jet energies are corrected to the hadronic scale
\cite{springerlink:10.1140/epjc/s10052-010-1512-2,Aad:2011he}. Jet candidates are
rejected in regions of the calorimeter where the jet energy is not yet
measured in an optimal way. Candidates consistent with spurious
calorimeter noise or energy spikes are also rejected.

Photon candidates are reconstructed from clusters in the
electromagnetic calorimeter and tracking information provided by the
inner detector. They satisfy standard ATLAS selection criteria that
are designed to reject instrumental backgrounds from hadrons
\cite{Aad:2011tw}. The photon candidates must meet $\pt$- and
$\eta$-dependent requirements on hadronic leakage, shower shapes in
the electromagnetic strip layer, and shower shapes in the second
sampling layer of the electromagnetic calorimeter. Inner detector
tracking information is used to reject electrons and to recover
photons converted to $e^+e^-$ pairs. Energy calibrations are applied
to photon candidates to account for energy loss in front of the
electromagnetic calorimeter and for both lateral and longitudinal
leakage. Events are discarded if the leading photon appears in
calorimeter cells affected by noise bursts or transient hardware
problems.

These photon identification criteria reduce instrumental backgrounds
to a negligible level, but much of the substantial background from
secondary (jet fragmentation) photons remains. We reduce this
background with requirements on nearby calorimeter
activity. Associated ``isolation'' calorimeter energy near the photon candidate
is calculated by summing the transverse momentum as measured in
electromagnetic and hadronic calorimeter cells inside a cone of radius
$\Delta R = \sqrt{ (\Delta \eta)^2 + (\Delta \phi)^2 } = 0.4$ centered
on the photon cluster, but excluding the energy of the photon cluster
itself. The isolation energy is corrected on an event-by-event basis
for the ambient energy density due to pile-up and the underlying
event. This isolation energy is required to be less than 7~\gev.

The photon deposits energy in the electromagnetic calorimeter in such
a way as also to be reconstructed as a jet. Jets within
$\Delta R < 0.2$ of the photon are therefore not considered in this analysis.
We require an angular separation $\Delta R(\gamma, \text{jet}) > 0.6$
between the signal photon and all jets with \pt $>$ 30~\gev to reduce the
background from photons during fragmentation of final state quarks
(fragmentation photons) and to reduce the systematic effects from
leakage of nearby jet showers into the photon isolation energy
measurement.

Additional reduction of fragmentation photon background is achieved by
requirements on the photon and jet pseudorapidities.
Dijet production rates increase
with jet pseudorapidity whereas rates for our assumed $s$-channel
signal would diminish. We restrict the analysis to photons in the
barrel calorimeter, $|\eta| < 1.37$, and require $|\eta_{\gamma} -
\eta_{j}| < 1.4$ between the photon and jet. The former criterion was
chosen to avoid kinematic bias of the \Mgj distribution due to
inclusion of any $\eta$ range where reconstruction efficiency is lower,
such as the barrel-endcap transition region $1.37 < |\eta| <
1.52$. The latter was chosen by optimizing expected significance using
the $|\eta_{\gamma} - \eta_{j}|$ distributions found in excited-quark
signal simulation and background-dominated control data selected as in
the nominal analysis but inverting the photon isolation
requirement. This control sample is also used to check the background
estimate.

\begin{figure}[!htb]
  \includegraphics[width=\columnwidth]{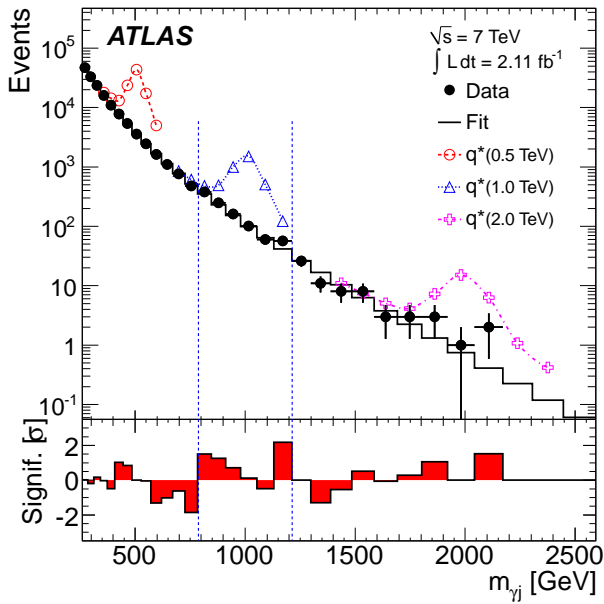}
  \caption{Invariant mass of the photon+jet pair for events passing
    the final selection. Overlaid: the fitted background function
    integrated over each bin (stepped solid line), the most discrepant
    region identified by \bumphunter (two dashed vertical lines), and
    three examples of excited-quark signals, normalized to luminosity,
    as described in the text. The bottom panel shows the statistical
    significance of the difference between data and background in each
    bin.}\label{fig:mgj}
\end{figure}

After the above selections, Fig.~\ref{fig:mgj} shows the distribution
of the \Mgj invariant mass in bins equal to the mass resolution. The
\Mgj resolution is about 4\% at 600~\gev, improving to 3\% at
2 \tev. We determine the combined SM and instrumental background to
the search by fitting this distribution to the four-parameter
{\it ansatz}
\begin{equation}\label{eq:bkgfunc}
f(x \equiv \Mgj/\sqrt{s} ) = p_1 (1-x)^{p_2} x^{-p_3 - p_4 \ln x}.
\end{equation}
\noindent The motivation for this function is discussed in Ref.~\cite{PhysRevLett.72.3004,PhysRevD.79.112002,PhysRevLett.105.161801,PhysRevLett.105.211801}.
The fit result is also shown in Fig.~\ref{fig:mgj}. The bottom panel 
of the figure shows the statistical significance of the difference
between data and the fit in each bin~\cite{2011arXiv1111.2062}.
With a negative
log-likelihood test statistic, the $p$-value is 23\%, indicating the
data distribution is compatible with Eq.~\ref{eq:bkgfunc}. The
functional form also describes the leading-order \pythia direct
photon prediction for comparable event statistics.

We search for statistical evidence of a resonance in this distribution
using the \bumphunter algorithm \cite{2011arXiv1101.0390C}. The
algorithm operates on the binned \Mgj distribution, comparing the
background estimate with the data in mass intervals of varying
contiguous bin multiplicities across the entire distribution. For each
interval in the scan, it computes the significance of any excess
found. The algorithm identifies the interval 784--1212~\gev, indicated
by the vertical lines in Fig.~\ref{fig:mgj}, as the single most
discrepant interval. The significance of the outcome is evaluated
using the ensemble of possible outcomes for the significance of any
region in the distribution in the background-only hypothesis, obtained
by repeating the analysis on pseudodata drawn from the background
function. Before including systematic uncertainties, the probability
($p$-value) of observing a background fluctuation at least as
significant as the above, including the trials factor, or
``look-elsewhere'' effect, is 20\%. Inclusion of systematic
uncertainties renders the $p$-value similarly large.

Lacking evidence of any signal, we exclude two types of photon+jet
signals: a generic signal with Gaussian distribution and arbitrary
production cross section, and the excited-quark model. We compute
Bayesian limits at 95\% credibility level (CL) using a prior
probability density that is constant for positive values of the signal
production cross section and zero for unphysical, negative values, as
described in Ref.~\cite{Aad:2011aj}. We consider systematic
uncertainties on expected signal yield due to imperfect knowledge of
the detector: the integrated luminosity (3.7\%), trigger efficiencies
($<$0.5\%), and signal photon identification efficiencies (2.0\%). The
last of these consists of isolation (0.4\%), pile-up interactions
(0.5\%), conversions (1.2\%), simulation mismodeling (1.3\%), and the
extrapolation of the photon identification efficiency to high \pt
($<$0.3\%). Uncertainties on photon energy scale (0.5--1.5\%), jet
energy scale (2--4\%), and jet energy resolution (5--15\%) contribute
through their effects on the signal distribution. These systematic
uncertainties are treated as marginalized Gaussian nuisance parameters
in the limit calculation.

We also evaluate two systematic uncertainties on the background
estimate. To account for the statistical uncertainties on the
background fit parameters, we repeatedly fit the background function
to pseudodata for each bin drawn from Poisson distributions. The mean
of the Poisson distribution for a given bin corresponds to the number
of entries actually observed in that bin in the data. We then take the
variation in the fit predictions for a given bin, 0.5\% of the
background at low mass to almost 10\% of the background at 2 TeV, as
indicative of the systematic uncertainty. We treat this bin-by-bin
uncertainty in the limit as fully correlated, using a single nuisance
parameter that scales the entire background distribution.

While our function can describe the \Mgj shape for direct photon
production, as modeled in the \pythia direct photon+jet simulation,
the function need not remain a good description of the full
distribution after including nonisolated and fragmentation photon
events. For example, the function describes the next-to-leading order
prediction implemented in \jetphox, which includes the fragmentation
photon contributions, for some viable choices of theory parameters but
not for others.

The second background systematic uncertainty accounts for any
unmodeled features of fragmentation photon events in our isolated
photon sample. We fit the background function to the \Mgj distribution
in the control data selected with the inverted isolation requirement,
then measure for each \Mgj bin the magnitude of any deviation from the
fit, and assign the ratio of the deviation to the fit expectation as
a parametrization bias systematic uncertainty. To extrapolate this
uncertainty to large \Mgj where few control data exist, we fit the
tail $\Mgj > 1 \tev$ with a two-degree polynomial.

\begin{figure}[!htb]
  \includegraphics[width=\columnwidth]{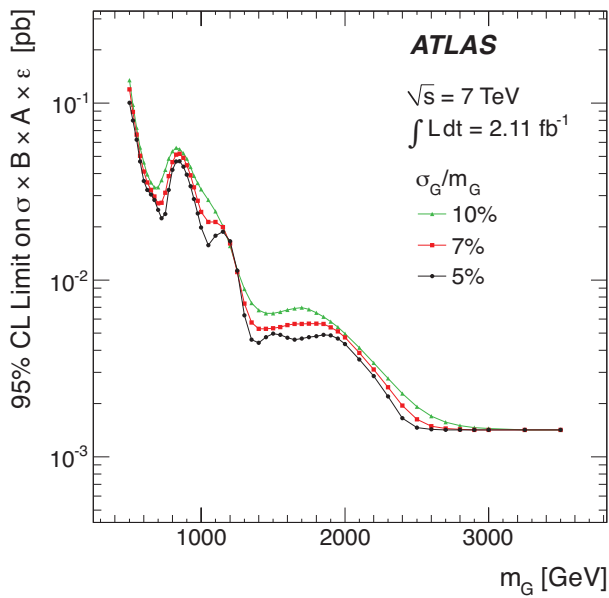}
   \caption{The 95\%\ CL upper limits on $\sigma \times \branching \times
  \acceptance \times \epsilon$ for a hypothetical signal with Gaussian \Mgj distribution
  decaying to a photon and a jet as a function of the signal mass, $m_{G}$,
  for three values of the relative Gaussian width $\sigma_G/m_{G}$
  and taking into account systematic
  uncertainties.}  \label{fig:gaussian-limit}
\end{figure}

Figure~\ref{fig:gaussian-limit} shows the model-independent limits on
the effective cross section, cross section $\sigma$ times branching
fraction \branching times acceptance \acceptance times efficiency
$\epsilon$, of a potential signal as a function of the central mass of
each signal template. We take the signal lineshape to be a Gaussian
distribution with one of three widths, $\sigma_G/m_{G} =$\xspace 5\%,
7\%, and 10\% of the central mass of the Gaussian. The limit weakens
as the width increases and the peak becomes less distinct. For
example, for a 1 \tev signal the limit for a width of 10\% is 1.6
times the limit for a width of 5\%.

\begin{figure}[!htb]
  \includegraphics[width=\columnwidth]{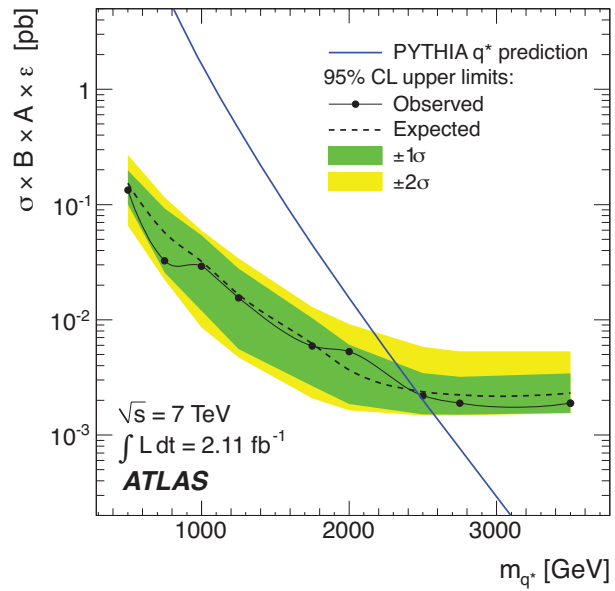}
  \caption{The 95\%\ CL upper limits on $\sigma \times \branching \times \acceptance \times \epsilon$ for
    excited quarks decaying to a photon and a jet as a function of the signal
    mass, $\Mqstar$, for the \pythia \qstar prediction  and taking into account systematic
    uncertainties.}
  \label{fig:excited-quark-limit} 
\end{figure}

The limit on the effective cross section in the excited-quark model is
shown in Fig.~\ref{fig:excited-quark-limit} as a function of the
\qstar mass. Also shown are $\pm1\sigma$ and $\pm2\sigma$ uncertainty
bands indicating the underlying distribution of possible limit
outcomes in the background-only hypothesis. The solid line indicates
the prediction from the \pythia excited-quark implementation. We
exclude such excited quarks with masses below 2.46 \tev at 95\% CL,
complementing the more stringent exclusion below 2.99 \tev on this
specific \qstar model in the dijet final state \cite{Aad:2011fq}.

In conclusion, the photon+jet mass distribution measured in
2.11~\ifb\xspace of $pp$ collision data collected at $\sqrt{s} =
7\;\tev$ by the ATLAS Collaboration has been examined for narrow
resonances. The observed distribution extends up to masses of about
2~\tev. It is well described by a smooth function fitted to it and
assumed to represent the SM expectation. No evidence for the
production of resonances is found. We set limits at 95\% CL on
Gaussian-lineshape and excited-quark signal using Bayesian
statistics. The limits on Gaussian resonances, for example, exclude
2~\tev resonances with effective cross sections near $5\;\text{\rm
  fb}$. We also exclude the excited-quark model in the photon+jet
final state for masses up to 2.46~\tev. The limits reported here on
resonant production of new particles in the photon+jet final state are
the most stringent limits set to date in this channel.

\bigskip
{\bf Acknowledgements}
\medskip

We thank CERN for the very successful operation of the LHC, as well as the
support staff from our institutions without whom ATLAS could not be
operated efficiently.

We acknowledge the support of ANPCyT, Argentina; YerPhI, Armenia; ARC,
Australia; BMWF, Austria; ANAS, Azerbaijan; SSTC, Belarus; CNPq and FAPESP,
Brazil; NSERC, NRC and CFI, Canada; CERN; CONICYT, Chile; CAS, MOST and
NSFC, China; COLCIENCIAS, Colombia; MSMT CR, MPO CR and VSC CR, Czech
Republic; DNRF, DNSRC and Lundbeck Foundation, Denmark; ARTEMIS, European
Union; IN2P3-CNRS, CEA-DSM/IRFU, France; GNAS, Georgia; BMBF, DFG, HGF, MPG
and AvH Foundation, Germany; GSRT, Greece; ISF, MINERVA, GIF, DIP and
Benoziyo Center, Israel; INFN, Italy; MEXT and JSPS, Japan; CNRST, Morocco;
FOM and NWO, Netherlands; RCN, Norway; MNiSW, Poland; GRICES and FCT,
Portugal; MERYS (MECTS), Romania; MES of Russia and ROSATOM, Russian
Federation; JINR; MSTD, Serbia; MSSR, Slovakia; ARRS and MVZT, Slovenia;
DST/NRF, South Africa; MICINN, Spain; SRC and Wallenberg Foundation,
Sweden; SER, SNSF and Cantons of Bern and Geneva, Switzerland; NSC, Taiwan;
TAEK, Turkey; STFC, the Royal Society and Leverhulme Trust, United Kingdom;
DOE and NSF, United States of America.

The crucial computing support from all WLCG partners is acknowledged
gratefully, in particular from CERN and the ATLAS Tier-1 facilities at
TRIUMF (Canada), NDGF (Denmark, Norway, Sweden), CC-IN2P3 (France),
KIT/GridKA (Germany), INFN-CNAF (Italy), NL-T1 (Netherlands), PIC
(Spain), ASGC (Taiwan), RAL (UK) and BNL (USA) and in the Tier-2
facilities worldwide.

\bibliography{letter}

\begin{thebibliography}{10}
\expandafter\ifx\csname urlstyle\endcsname\relax
  \providecommand{\doi}[1]{doi:\discretionary{}{}{}#1}\else
  \providecommand{\doi}{doi:\discretionary{}{}{}\begingroup
  \urlstyle{rm}\Url}\fi

\bibitem{Catani:2002ny}
S.~Catani, M.~Fontannaz, J.~Guillet, and E.~Pilon, \href{http://dx.doi.org/10.1088/1126-6708/2002/05/028}{JHEP \textbf{0205}, 028 (2002).}

\bibitem{Aurenche:2006vj}
P.~Aurenche, M.~Fontannaz, J.-P. Guillet, E.~Pilon, and M.~Werlen, \href{http://dx.doi.org/10.1103/PhysRevD.73.094007}{Phys. Rev. D
  \textbf{73}, 094007 (2006).
}
\bibitem{Belghobsi:2009hx}
Z.~Belghobsi {\it et~al.}, \href{http://dx.doi.org/10.1103/PhysRevD.79.114024}{Phys. Rev. D \textbf{79}, 114024 (2009).
}
\bibitem{PhysRevD.82.014015}
R.~Ichou and D.~d'Enterria, \href{http://dx.doi.org/10.1103/PhysRevD.82.014015}{Phys. Rev. D \textbf{82}, 014015 (2010).
}
\bibitem{Note0} The invariant mass \Mgj is defined as
  $\sqrt{(E^{\gamma} + E^{j})^2 - (\vec{p}^{\gamma} +
    \vec{p}^{j})^2}$, where $E$ and $\vec{p}$ denote the
  energy and momentum, respectively, of the photon and the jet.
\bibitem{Bhattacharya:2009p2168}
S.~Bhattacharya, S.~S. Chauhan, B.~C. Choudhary, and D.~Choudhury, \href{http://dx.doi.org/10.1103/PhysRevD.80.015014}{Phys. Rev. D
  \textbf{80}, 015014 (2009).
}
\bibitem{PhysRevD42815}
U.~Baur, M.~Spira, and P.~M. Zerwas, \href{http://dx.doi.org/10.1103/PhysRevD.42.815}{Phys. Rev. D \textbf{42}, 815 (1990).
}
\bibitem{kang-and-luty}
J.~Kang and M.~A. Luty, \href{http://dx.doi.org/10.1088/1126-6708/2009/11/065}{
JHEP \textbf{11}, 065 (2009).
}
\bibitem{PhysRevD.83.035019}
S.~P. Martin, \href{http://dx.doi.org/10.1103/PhysRevD.83.035019}{
Phys. Rev. D \textbf{83}, 035019 (2011).
}
\bibitem{Okun19801}
L.~B. Okun, \href{http://dx.doi.org/10.1016/0550-3213(80)90439-3}{
Nucl. Phys. \textbf{B173}, 1 (1980).
}
\bibitem{PhysRevD.78.016005}
L.~A. Anchordoqui, H.~Goldberg, S.~Nawata, and T.~R. Taylor, \href{http://dx.doi.org/10.1103/PhysRevD.78.016005}{Phys. Rev. D \textbf{78}, 016005 (2008).}

\bibitem{PhysRevD.60.034004}
O.~\ifmmode \mbox{\c{C}}\else \c{C}\fi{}ak\ifmmode \imath \else \i \fi{}r and R.~Mehdiyev, \href{http://dx.doi.org/10.1103/PhysRevD.60.034004}{Phys. Rev. D \textbf{60}, 034004 (1999).
}

\bibitem{PhysRevLett.101.241803}
L.~A. Anchordoqui {\it et~al.}, \href{http://dx.doi.org/10.1103/PhysRevLett.101.241803}{Phys. Rev. Lett. \textbf{101}, 241803 (2008).
}

\bibitem{Nath2010185}
P.~Nath {\it et~al.}, \href{http://dx.doi.org/DOI: 10.1016/j.nuclphysbps.2010.03.001}{
Nucl. Phys. B - Proceedings Supplements \textbf{200-202}, 185 (2010).
}

\bibitem{Bai2010292}
Y.~Bai and A.~Martin, \href{http://dx.doi.org/10.1016/j.physletb.2010.08.058}{
Phys. Lett. B \textbf{693}, 292 (2010).
}

\bibitem{PhysRevLett.72.3004}
F.~Abe {\it et~al.}
\newblock ({{CDF Collaboration}}), \href{http://dx.doi.org/10.1103/PhysRevLett.72.3004}{
Phys. Rev. Lett. \textbf{72}, 3004 (1994).
}

\bibitem{adloff-2000-17} C.~Adloff {\it et~al.}  \newblock ({{H1
    Collaboration}}),
  \href{http://dx.doi.org/10.1007/s100520000503}{Eur. Phys. J. C
    \textbf{17}, 567 (2000).  }

\bibitem{zeus-collaboration-2002-549} S.~Chekanov {\it et~al.}
  \newblock ({{ZEUS Collaboration}}),
  \href{http://dx.doi.org/doi:10.1016/S0370-2693(02)02863-0}{
    Phys. Lett. B \textbf{549}, 32 (2002).  }

\bibitem{PhysRevD-65-052006}
T.~Affolder {\it et~al.}
\newblock ({{CDF Collaboration}}), \href{http://dx.doi.org/10.1103/PhysRevD.65.052006}{
Phys. Rev. D \textbf{65}, 052006 (2002).
}

\bibitem{Aad:2011tw}
ATLAS Collaboration, \href{http://dx.doi.org/10.1016/j.physletb.2011.11.010}{Phys. Lett. B \textbf{706}, 150 (2011).}

\bibitem{Chatrchyan:2011ue}
CMS Collaboration, \href{http://dx.doi.org/10.1103/PhysRevD.84.052011}{Phys. Rev. D \textbf{84}, 052011 (2011)}.

\bibitem{Aad:2011fq}
ATLAS Collaboration, \href{http://dx.doi.org/10.1016/j.physletb.2012.01.035}{Phys. Lett. B \textbf{708}, 37 (2012)}.

\bibitem{Pythia:2001}
T.~Sjostrand {\it et~al.}, \href{http://dx.doi.org/10.1016/S0010-4655(00)00236-8}{
Comput. Phys. Commun. \textbf{135}, 238 (2001)}.

\bibitem{ATLAS-CONF-2010-031}
ATLAS Collaboration, Report No. \href{http://cdsweb.cern.ch/record/1277665}{ATLAS-CONF-2010-031} (unpublished).
%http://cdsweb.cern.ch/record/1266235

\bibitem{Sherstnev:2007nd}
A. Sherstnev and R. S. Thorne, \href{http://dx.doi.org/10.1140/epjc/s10052-008-0610-x}{Eur. Phys. J. C \textbf{55}, 553 (2008).}

\bibitem{geant}
S. Agostinelli {\it et~al.} (GEANT4 Collaboration), \href{http://dx.doi.org/10.1016/S0168-9002(03)01368-8}{Nucl. Instrum. Methods in Phys. Res. Sect. A {\textbf{A506}}, 250 (2003)}.

\bibitem{mcsim}
ATLAS Collaboration, \href{http://dx.doi.org/10.1140/epjc/s10052-010-1429-9}{Eur. Phys. J. C {\textbf{70}}, 823 (2010)}.


\bibitem{1748-0221-3-08-S08003}
ATLAS Collaboration, \href{http://dx.doi.org/10.1088/1748-0221/3/08/S08003}{JINST \textbf{3}, S08003 (2008)}.

\bibitem{Note1}
ATLAS uses a right-handed coordinate system with the $z$-axis along the beam
  pipe. The $x$-axis points to the centre of the LHC ring, and the $y$-axis
  points upward. Cylindrical coordinates ($r$,$\phi $) are used in the
  transverse plane, with azimuthal angle $\phi $. The pseudorapidity is defined
  in terms of the polar angle $\theta $ as $\protect \ensuremath {\etaa
  }\protect \xspace \equiv \protect \mathrm {-ln\ tan}(\theta /2)$. Transverse
  momentum and energy are defined as $\pt = p$\protect \,sin\,$\theta $ and $\et
  = E$\protect \,sin\,$\theta $, respectively.

%% \footnote{ATLAS uses a right-handed coordinate system with
%% the $z$-axis along the beam pipe. The $x$-axis points to the centre of
%% the LHC ring, and the $y$-axis points upward. Cylindrical coordinates
%% ($r$,$\phi$) are used in the transverse plane, with azimuthal angle
%% $\phi$.  The pseudorapidity is defined in terms of the polar angle
%% $\theta$ as $\eta \equiv \mathrm{-ln\ tan}(\theta/2)$. Transverse
%% momentum and energy are defined as $p_T = p$\,sin$\theta$ and $E_T =
%% E$\,sin$\theta$, respectively.}

\bibitem{springerlink:10.1140/epjc/s10052-010-1512-2}
ATLAS Collaboration, \href{http://dx.doi.org/10.1140/epjc/s10052-010-1512-2}{Eur. Phys. J. C \textbf{71}, 1512 (2011)}.

\bibitem{Aad:2011he}
ATLAS Collaboration, \href{http://arxiv.org/abs/1112.6426}{arXiv:1112.6426 [Eur. Phys. J. C (to be published)]}.
%       key            = "1082939",
%       title          = "{Jet energy measurement with the ATLAS detector in
%                         proton-proton collisions at sqrt(s) = 7 TeV}",
%       collaboration  = "ATLAS Collaboration",
%       year           = "2011",
%       eprint         = "1112.6426",
%       archivePrefix  = "arXiv",
%       primaryClass   = "hep-ex",
%       reportNumber   = "CERN-PH-EP-2011-191",
%       SLACcitation   = "%%CITATION = ARXIV:1112.6426;%%",
% }
\bibitem{ATLAS-CONF-2011-116}
ATLAS Collaboration, Report No. \href{http://cdsweb.cern.ch/record/1376384}{ATLAS-CONF-2011-116} (unpublished).

% luminosity
\bibitem{Aad:2011dr}
ATLAS Collaboration, \href{http://dx.doi.org/10.1140/epjc/s10052-011-1630-5}{Eur. Phys. J. C \textbf{71}, 1630 (2011).}

%% \bibitem{2010EPJC...70..723A}
%% ATLAS Collaboration, \href{http://dx.doi.org/10.1140/epjc/s10052-010-1354-y}{Eur. Phys. J. C \textbf{70}, 723 (2010)}.

%% \bibitem{2010EPJC...70.1193A}
%% ATLAS Collaboration, \href{ohttp://dx.doi.org/10.1140/epjc/s10052-010-1508-y}{Eur. Phys. J. C \textbf{70}, 1193 (2010)}.

\bibitem{Cacciari:2008gp}
M.~Cacciari, G.~P. Salam, and G.~Soyez, \href{http://dx.doi.org/10.1088/1126-6708/2008/04/063}{JHEP \textbf{0804}, 063 (2008).
}

\bibitem{PhysRevD.79.112002}
T.~Aaltonen {\it et~al.}
\newblock ({{CDF Collaboration}}), \href{http://dx.doi.org/10.1103/PhysRevD.79.112002}{Phys. Rev. D \textbf{79}, 112002 (2009)}.

\bibitem{PhysRevLett.105.161801}
ATLAS Collaboration, \href{http://dx.doi.org/10.1103/PhysRevLett.105.161801}{Phys. Rev. Lett. \textbf{105}, 161801 (2010).}

\bibitem{PhysRevLett.105.211801}
CMS Collaboration, \href{http://dx.doi.org/10.1103/PhysRevLett.105.211801}{Phys. Rev. Lett. \textbf{105}, 211801 (2010).}

\bibitem{2011arXiv1111.2062}
G.~Choudalakis and D.~Casadei, \href{http://arxiv.org/abs/1111.2062}{arXiv:1111.2062} (unpublished).

\bibitem{2011arXiv1101.0390C}
G.~Choudalakis, \href{http://arxiv.org/abs/1101.0390}{arXiv:1101.0390} (unpublished).

\bibitem{Aad:2011aj}
ATLAS Collaboration, \href{http://dx.doi.org/10.1088/1367-2630/13/5/053044}{New J. Phys. \textbf{13}, 053044 (2011)}.

%% \href{http://arxiv.org/abs/1103.3864}{arXiv:1103.3864}
%%   {ATLAS$\;$Collaboration}(2011{\natexlab{a}})}]{Aad:2011aj}%
%%   \BibitemOpen
%%   \bibfield  {author} {\bibinfo {author} {\bibnamefont
%%   {ATLAS$\;$Collaboration}},\ }\href@noop {} {\bibfield  {journal} {\bibinfo
%%   {journal} {New Journal of Physics}\ }\textbf {\bibinfo {volume} {13}},\
%%   \bibinfo {pages} {053044} (\bibinfo {year} {2011}{\natexlab{a}})},\ \Eprint
%%   {http://arxiv.org/abs/1103.3864} {arXiv:1103.3864 [hep-ex]} \BibitemShut
%%   {NoStop}%

%% \bibitem{Warsinsky:1290569}
%% M.~Warsinsky {\it et~al.}  ATL-PHYS-INT-2010-101 (2010).


\end{thebibliography}
\onecolumngrid

\clearpage 
\onecolumngrid
% ATLAS Collaboration author list for 31-OCT-2011
% Data extracted on 16-Mar-2012 for paperid 150
%\documentclass[11pt]{article}
%\usepackage{a4wide}\begin{document}
\begin{flushleft}
{\Large The ATLAS Collaboration}

\bigskip

G.~Aad$^{\rm 48}$,
B.~Abbott$^{\rm 110}$,
J.~Abdallah$^{\rm 11}$,
A.A.~Abdelalim$^{\rm 49}$,
A.~Abdesselam$^{\rm 117}$,
O.~Abdinov$^{\rm 10}$,
B.~Abi$^{\rm 111}$,
M.~Abolins$^{\rm 87}$,
H.~Abramowicz$^{\rm 152}$,
H.~Abreu$^{\rm 114}$
% ...
\bigskip

% ...
$^{10}$ Institute of Physics, Azerbaijan Academy of Sciences, Baku, Azerbaijan\\
$^{11}$ Institut de F\'isica d'Altes Energies and Departament de F\'isica de la Universitat Aut\`onoma  de Barcelona and ICREA, Barcelona, Spain\\
% ...
$^{48}$ Fakult\"{a}t f\"{u}r Mathematik und Physik, Albert-Ludwigs-Universit\"{a}t, Freiburg i.Br., Germany\\
$^{49}$ Section de Physique, Universit\'e de Gen\`eve, Geneva, Switzerland\\
% ...
$^{87}$ Department of Physics and Astronomy, Michigan State University, East Lansing MI, United States of America\\
% ...
$^{110}$ Homer L. Dodge Department of Physics and Astronomy, University of Oklahoma, Norman OK, United States of America\\
$^{111}$ Department of Physics, Oklahoma State University, Stillwater OK, United States of America\\
% ...
$^{114}$ LAL, Univ. Paris-Sud and CNRS/IN2P3, Orsay, France\\
% ...
$^{117}$ Department of Physics, Oxford University, Oxford, United Kingdom\\
% ...
$^{152}$ Raymond and Beverly Sackler School of Physics and Astronomy, Tel Aviv University, Tel Aviv, Israel\\
% ...
\end{flushleft}

%\end{document} 

\end{document}